\documentclass[12pt,fleqn]{article}
\usepackage{amsmath}
\usepackage{cite}
\usepackage{amsfonts}
\newlength{\dinwidth}
\newlength{\dinmargin}
\begin{document}

\begin{flushright}
DFF302/0398 
\end{flushright}
\vspace{1cm}

\begin{center}

{\large{\textbf{Energy Scale(s) and Next-to-leading BFKL Equation}
\footnote{Work supported in part by EU Network contract FMRX-CT98-0194 and by
M.U.R.S.T. (Italy)}}}

\vspace{1cm}

\normalsize{Marcello Ciafaloni and Gianni Camici} 
 \vspace{1cm}

\textit{Dipartimento di Fisica dell'Universit\`a, Firenze\\
and INFN, Sezione di Firenze, Italy} 

\end{center}
\vspace{1cm}

\begin{abstract}
We complete the calculation of the next-to-leading kernel of the BFKL equation by 
disentangling its energy-scale dependent part from the impact factor corrections
in large-
\textbf{k} dijet production. Using the irreducible part previously obtained,
we derive the final form of the kernel eigenvalue and of the hard Pomeron shift
for various scales. We also discuss the scale changes, the physical equivalence of
a class of scales, and how to use the collinear safe ones. 
\end{abstract}
\vspace{1cm}
\begin{center}
\end{center}

\newpage

The theoretical effort devoted in the past few years \cite{1,2,3,4,5,6,7,8,9,10} 
to the next-to-leading (NL) corrections to the BFKL equation \cite{1}, 
is now approaching 
its final steps \cite{11,12}.

After the calculation of the relevant high-energy vertices \cite{2,3,4,5,6,7} 
by Fadin, Lipatov and other 
authors, and of the $q\bar{q}$ part of the kernel \cite{8,10} by the 
present authors,
we computed the eigenvalue of the ``irreducible'' part of the gluonic kernel
\cite{9}, by pointing out that the left-over ``leading terms'', 
after subtraction of a common scale,
would lead to an additional contribution to the kernel, yet to be determined.

The purpose of this note is to complete the calculation above with its missing 
part, and to discuss
 the energy-scale dependence of the kernel. 
Our results, which refer to large-$\mathbf{k}$ dijet production, are based on the separation of the NL kernel 
from the one-loop partonic impact factors, recently determined by one of us \cite{11}, and 
depending on some input scale $s_{0}$.
For the case $s_{0} = k_{1}k_{2}$ (where the $\mathbf{k}$'s are the 
transverse momenta of the jets), the full kernel derived
here agrees with the one recently proposed  
by Fadin and Lipatov \cite{12},  who do not discuss the impact factors'
contribution.

The starting point of our argument is a \textbf{k}-factorized form of dijet
production in $a-b$ parton scattering of the type introduced by one of us \cite{11}
\begin{eqnarray}
\frac{d\sigma_{ab}}{d[\mathbf{k}_1]d[\mathbf{k}_2]}
&\equiv&
h_a^{(0)}(\mathbf{k}_{1}) h_{b}^{(0)}(\mathbf{k}_{2}) \rho_{ab}(\mathbf{k}_{1},
\mathbf{k}_{2})=\\
&=&
\int\frac{d\omega}{2\pi i}\left(\frac{s}{s_{0}(\mathbf{k}_{1},\mathbf{k}_{2})}
\right)
^{\omega}\frac{\pi}{\omega} h_{a}(\mathbf{k}_{1}) {\cal G}_{\omega}(\mathbf{k}_{1}, 
\mathbf{k}_{2})
h_{b}(\mathbf{k}_{2}),\nonumber
\label{1}
\end{eqnarray}
where $d[\mathbf{k}] = d^{2(1+\epsilon)}\mathbf{k}/\pi^{1+\epsilon}$ in $D=4+2\epsilon$ 
dimensions, $h_a(a=q,g)$ are the partonic impact factors, 
which at Born level are of
type $ h^{(0)}_a\sim C_a\alpha_s/\mathbf{k}^2\,\, (C_a = C_F,C_A)$, 
and ${\cal G}_{\omega}$
is the gluon Green's function, given by
\begin{equation}
{\cal G}_{\omega} = (1+\overline{\alpha}_s H_L) \;
G_{\omega}(1+\overline{\alpha}_s H_R)
\label{2}
\end{equation}
where $H_L,H_R$ are $\omega$-independent kernels to be defined below, and the resolvent
\begin{equation}
G_{\omega} =[1-\frac{\overline\alpha_s}{\omega}(K_0+K_{NL})]^{-1},\hspace{2cm}
\overline{\alpha}_s
= \frac{\alpha_s N_C}{\pi},
\label{3}
\end{equation}
is given in terms of the leading kernel $K_{0}$ and of the NL one $K_{NL}$,
whose scale-invariant part \cite{9} we denote by $\alpha_{s} K_{1}$.

The representation (1) differs from the one adopted in \cite{11} by the introduction
in the Green's function (2) of the $NL$ operator factors $H_{L},H_{R}$ which,
for a class of scale choices, cannot be incorporated in a redefinition of the
impact factors (see below). Both the impact factors and the $NL$ kernel are in
general dependent on the scale $s_{0}$ for which the representation (1) is
assumed. It is presumed that such dependence will not lead to physical differences
in observable quantities.

In order to understand how scale, impact factors and kernel are tight-up
together, let us expand the r.h.s. of Eq~(1) up to two-loop level,
by setting $H_{L}=H_{R}=0$ for simplicity. One has
\begin{eqnarray}
\overline{\alpha}_{s} (h^{(0)}_{a}h^{(0)}_{b} K_{0} \log \frac{s}{s_{0}}
	&+&
h^{(0)}_{a}h^{(1)}_{b} + h^{(1)}_{a}h^{(0)}_{b}) + 
\overline{\alpha}_{s} 
\left[ \frac{1}{2} 
\left( \log \frac{s}{s_{0}}
\right)^{2} \cdot h^{(0)}_{a} \overline{\alpha}_{s}K_{0}^{2} h^{(0)}_{b} + 
\right. \\
 	+ \log \frac{s}{s_{0}} (h^{(0)}_{a}\overline{\alpha}_{s}K_{0}h^{(1)}_{b} 
&+& 
\left.
	h^{(1)}_{a}\overline{\alpha}_{s}K_{0}h^{(0)}_{b} + 
h^{(0)}_{a}h^{(0)}_{b} \alpha_{s}K_{1})
	+ \textrm{const} 
\right] + \dots \; , \nonumber
\label{4}
\end{eqnarray}
where the superscripts refer to the order of the expansion. A similar
expansion holds, for say $H_{R} \neq 0$, with $h^{(1)}_{b}$ replaced
by $h^{(1)}_{b} + H_{R} h^{(0)}_{b}$.

From the one-loop terms in round brackets of Eq.~(4) one can determine
the leading kernel $K_{0}$ \cite{1} unambiguously, as the coefficient
of log $s$. However, at $NL$ order, the impact factor $h^{(1)}$ can
only be determined \cite{11} after subtracting the log $s_{0}$ contribution,
and is thus $s_{0}$-dependent.

Similarly, the $NL$ kernel can be extracted from the two-loop 
contribution in square brackets only after the one-loop impact
factors have been found, and is thus $s_{0}$-dependent too.
For this reason the derivation in Ref.~\cite{12}, which is not
supplemented by a determination of the one-loop impact factors,
appears to be incomplete.

The one-loop analysis of Eq.~(4) was carried out in Ref.~\cite{11}.
The problem was to match the central region with the fragmentation
region in the squared matrix element for emitting a gluon of
transverse momentum $\mathbf{q}=\mathbf{k}_{1}-\mathbf{k}_{2}$ and 
momentum fraction $z_{1}=qe^{y}/\sqrt{s}$. By explicit calculation
of the fragmentation vertex $F_{0}(z_{1})$, it was found that,
for initial quarks,
\begin{eqnarray}
\rho^{(1)}_{1g} (\mathbf{k}_{1},\mathbf{k}_{2})
	&=& \frac{\overline{\alpha}_{s}}{\mathbf{q}^{2}} \left(
\int^{1}_{\mathbf{q}/\sqrt{s}} \frac{dz_{1}}{z_{1}} F_{0} (z_{1}, \mathbf{k}_{1},
\mathbf{q}) + (1\leftrightarrow 2)\right) \nonumber \\
	&=& \frac{\overline{\alpha}_{s}}{\mathbf{q}^{2}} \left(
\log \frac{\sqrt{s}}{\textrm{Max}(q,k_{1})} + h_{q} (\mathbf{q},
 \mathbf{k}_{1},) + (1\leftrightarrow 2)\right), \\
	&=& \frac{\overline{\alpha}_{s}}{\mathbf{q}^{2}} \left(
\log \frac{\sqrt{s}}{k_{1}} - \log\frac{q}{k_{1}} 
\Theta_{qk_{1}} + h_{q} (\mathbf{q},
 \mathbf{k}_{1}) + (1\leftrightarrow 2)\right) \nonumber 
\label{5}
\end{eqnarray}
where the additional constant $h_{q}$ \cite{11} is characterized
by its vanishing at $q=0$ and by its collinear limit at $k_{1}=0$,
related to the non singular part $\hat{\gamma}_{gq} (\omega =0)$
of the anomalous dimension matrix element.

The physical content of Eq~(5) is that the gluon phase space
$0<y<Y_{q}\equiv \log\frac{\sqrt{s}}{q}$ (available in full for
$q>k_{1}$), is dynamically cut-off by the fragmentation vertex
for $q<k_{1}$, yielding the integration region $0<y<Y_{1}\equiv
\log\frac{\sqrt{s}}{k_{1}}$, which corresponds to the angular
ordering of the emitted gluon (coherence effect \cite{13}).

Mathematically, Eq~(5) implies, for the scale $s_{>}=$ Max $(k_{1},q)
\cdot$ Max $(k_{2},q)$, that the $\mathbf{k}_{1}$-integral of the
constant piece $h_{q}/\mathbf{q}^{2}$ yields the one-loop correction to the
quark impact factor \cite{11}
\begin{equation}
\left. h^{(1)}_{q}/h^{(0)}_{q}\right)_{\textrm{real}} =
\overline{\alpha}_{s}(\mathbf{k}^{2})^{\epsilon}
\left[ \left(\frac{1}{4}-\frac{3}{4\epsilon}\right) -
\frac{\pi^{2}}{3}-\frac{1}{2} \right] \; ,
\label{6}
\end{equation}
with the expected collinear singularity due to 
$\hat{\gamma}_{gq} (\omega =0)$. For this reason one could set
$H_{L}=H_{R}=0$ in this case,
and the scale $s_{>}$ was singled out as collinear safe.

For other scale choices, the result in Eq~(5) implies
$H_{L},H_{R} \neq 0$ in general. For instance, if we take
$s_{0}=k_{1}k_{2}$, the 3${}^{rd}$ line of Eq~(5) shows
some additional constant contributions, of the form
\begin{eqnarray}
\overline{\alpha}_{s} H = \overline{\alpha}_{s} H_{R}
(\mathbf{k}_{1},\mathbf{k}_{2}) 
&\equiv&
- \frac{\overline{\alpha}_{s}}{\mathbf{q}^{2}}
\log\frac{q}{k_{2}} \Theta_{qk_{2}} \equiv
- \frac{\overline{\alpha}_{s}}{\mathbf{q}^{2}}
l(q,k_{2}) \nonumber \\
\left(H_{L} (\mathbf{k}_{1},\mathbf{k}_{2}) \right.
&=& \left. H_{R}
(\mathbf{k}_{2},\mathbf{k}_{1}) = H^{\dag} \right) \; ,
\label{7}
\end{eqnarray}
which \textit{cannot} be integrated over $\mathbf{k}_{2} (\mathbf{k}_{1})$
because of their unacceptable collinear behaviour ($\frac{1}{\epsilon^{2}}$
poles). Therefore, the kernels $H$ and $H^{\dag}$ are to be incorporated
in the gluon Green's function, as anticipated in Eq.~(2). Their collinear
behaviour is apparent from the eigenvalues of $H(H^{\dag})$ on test functions
$\sim(\mathbf{k}^{2})^{\gamma-1}$ , which are given by
$h(\gamma) (h(1-\gamma))$, where
\begin{equation}
h(\gamma) = -\frac{1}{2} \sum^{\infty}_{n=0} \left(
\frac{\Gamma(\gamma+n)}{\Gamma(\gamma)n!}\right)^{2} \;
\frac{1}{(\gamma+n)^{2}} \stackrel{\gamma\rightarrow 0}{=}
-\frac{1}{2\gamma^{2}} +O(\gamma^{2}) \; ,
\label{8}
\end{equation}
thus showing a $1/\gamma^{2}(1/(1-\gamma)^{2})$ singularity.

For the reasons above it was concluded that the scale $k_{1}k_{2}$
\textit{is not} collinear safe (while $s_{>}$ is), even
if a representation of the type (1) can be written in both
cases up to two loops, with properly chosen $H_{R}$ and $H_{L}$.

The real problem is, however, for which scale(s) the representation
in Eq~(1) is supposed to be valid \textit{to all orders}. This question
has no clearcut answer yet. However, the scales $k_{1}^{p}k_{2}^{2-p}$,
whose $\mathbf{k}$-dependence is factorized (like $k_{1}k_{2}, k_{1}^{2} ,
k_{2}^{2}$), are favoured by $t$-channel unitarity arguments \cite{14}
 and by the use of $O(2,1)$ group-theoretical variables \cite{15}. For
this reason we shall mostly consider this class of scales.

On the other hand, the scales of the type $s_{>}$ used before,
(e.g., $s_{M}=$Max $(\mathbf{k}_{1}^{2},\mathbf{k}_{2}^{2})$) are automatically
collinear safe, but are not factorized in their $\mathbf{k}$-dependence.
We shall comment on their use later on, because switching from one class
of scales to the other to all orders is not simply achieved by a
change in the two-loop $NL$ kernel.

Let us first derive the scale-dependent $NL$ corrections to the kernel
for the scale $k_{1}k_{2}$. The ``leading terms" at two-loop level
were defined in Ref~\cite{9} as
\begin{equation}
\rho^{(2)}_{1g} = `` \int dy \left[ \omega_{1}(Y_{1}+Y_{2}-2y) 
+ \omega_{2}(Y_{1}+Y_{2}+2y)\right] \;
\frac{\overline{\alpha}_{s}}{\mathbf{q}^{2}} \;"
\label{9}
\end{equation}
for one-gluon emission, and as
\begin{equation}
\rho^{(2)}_{2g} = `` \int dy_{1}dy_{2}
\frac{d[\mathbf{k}]\overline{\alpha}_{s}^{2}}{\mathbf{q}_{1}^{2}
\mathbf{q}_{2}^{2}} \;"
\label{10}
\end{equation}
for two-gluon emission, where
\begin{eqnarray}
\omega_{i} &=& \omega(\mathbf{k}^{2}_{i}) \; , \;
\omega(\mathbf{k}^{2}) = -\frac{\overline{\alpha}_{s}}{2\epsilon}\; 
\frac{\Gamma^{2}(1+\epsilon)}{\Gamma(1+2\epsilon)} (\mathbf{k}^{2})^{\epsilon}
\; ,\\
Y_{i} &=& \log \frac{\sqrt{s}}{k_{i}}\; , \; 
y_{i}=\log \left(z_{i}\frac{\sqrt{s}}{q_{i}}\right) \; , \;
\mathbf{q}_{i}^{2} = (\mathbf{k}_{i}-\mathbf{k})^{2} \;. \nonumber
\label{11}
\end{eqnarray}

The rapidity integrations in the r.h.s. of Eqs~(9) and (10) are meant
to be defined by matching with the fragmentation vertices. For instance
in the fragmentation region of $a$, Eq~(9) should be replaced,
at $NL$ level, by
\begin{eqnarray}
\rho^{(2)}_{1g} 
&=&  \frac{\overline{\alpha}_{s}}{\mathbf{q}^{2}} \int^{1}_{q/\sqrt{s}}
	 \left[ F_{0}(z_{1},\mathbf{k}_{1},\mathbf{q}) \omega_{2}
	\left(Y_{1}+Y_{2}+2 (Y_{q}+\log z_{1})\right) + \right. \\ 
&+&  \left. F_{1}(z_{1},\mathbf{k}_{1},\mathbf{q}) \omega_{1}
	\left(Y_{1}+Y_{2}-2 (Y_{q}+\log z_{1})\right)\right] \; , \nonumber
\label{12}
\end{eqnarray}
where $F_{0}$ is - by $\mathbf{k}$-factorization - the one-loop fragmentation
vertex occurring in Eq~(5). In principle, we should calculate the vertex $F_{1}$
also. However, at $NL$ level, we can set $F_{1}=1$ because the factor
in front of $F_{1}$ is just a constant in the fragmentation region of $a$,
so that $F_{1}-1$ contributes at $NNL$ level only.

We can thus calculate the logarithmic terms in $\rho^{(2)}_{1g} $ on the
basis of Eqs~(12) and (5). By simple algebra we find, after factorization of 
the one-loop impact factors of Eq~(6), the expression
\begin{eqnarray}
\frac{\partial\rho^{(2)}_{1g}}{\partial \log s}
=  \frac{\overline{\alpha}_{s}}{\Gamma(1-\epsilon)\mathbf{q}^{2}}
	\left[  \omega_{2}
	\left(\frac{1}{2} (Y_{1}+Y_{2})+2 (Y_{1}-l (q,k_{1}))\right)
+ \omega_{1} \frac{1}{2} (Y_{1}+Y_{2}) + 1 \leftrightarrow 2 \right] \\ 
=  \frac{\overline{\alpha}_{s}}{\Gamma(1-\epsilon)\mathbf{q}^{2}}
	\left[  2(\omega_{1}+\omega_{2})
	 (Y_{1}+Y_{2})-2\omega_{1} l (q,k_{2}) -  2\omega_{2} l(q,k_{1})+
	(Y_{1}-Y_{2}) (\omega_{1}-\omega_{2}) \right]\; , \nonumber
\label{13}
\end{eqnarray}
where $l(q,k_{1})$ occurs in the definition of $H$ and $H^{\dag}$ (Eq~(7)).

A similar calculation can be performed for matching the two gluon emission
density of Eq~(10) with the fragmentation regions. By $\mathbf{k}$-factorization,
the $F_{0}$ vertex only turns out to be relevant and, by using Eq~(5) and by
factorizing the one-loop impact factors, we obtain
\begin{equation}
\frac{\partial\rho^{(2)}_{2g}}{\partial \log s}
= \frac{\overline{\alpha}_{s}^{2}}{\Gamma^{2}(1-\epsilon)\mathbf{q}^{2}_{1}
	\mathbf{q}^{2}_{2}}
	 \left( Y_{1}+Y_{2}-l (q_{1},k_{1})-l(q_{2},k_{2})\right)\; .
\label{14}
\end{equation}

We recognize, in the sum of Eqs~(13) and (14), the logarithmic terms at scale
$k_{1}k_{2}$, due to the iteration of the leading kernel
\begin{equation}
\overline{\alpha}_{s} K_{0} = \frac{\overline{\alpha}_{s}}
{\mathbf{q}^{2}\Gamma(1-\epsilon)} +2\omega (\mathbf{k}^{2})\pi^{1+\epsilon}
\delta^{2(1+\epsilon)} (\mathbf{q})\, .
\label{15}
\end{equation}
By subtracting them out, we obtain the $NL$ part
\begin{eqnarray}
\left. \frac{d\sigma^{(2)}_{ab}}{d\log s} \right)_{NL} = \overline{\alpha}_{s}^{2}
\left[
h^{(0)}_{a} K_{0} (h^{(1)}_{b} + Hh^{(0)}_{b}) +
(h^{(1)}_{a} + h^{(0)}_{a}H^{+}) K_{0}h^{(0)}_{b}\right]-\\
h^{(0)}_{a}h^{(0)}_{b} \frac{\overline{\alpha}_{s}}{\mathbf{q}^{2}} \log\frac{k_{1}}{k_{2}}
(\omega_{2}-\omega_{1}), \nonumber
\label{16}
\end{eqnarray}

where $H$ is defined as in Eq~(7). By comparing with Eq~(4) (with $H\neq 0$),
we finally identify the scale-dependent contribution to the $NL$ kernel at scale
$k_{1}k_{2}$ as the last term in Eq.~(13), which, for $\epsilon \rightarrow 0$,
reads
\begin{equation}
\Delta (\alpha_{s}K_{1})= -\frac{1}{4} \overline{\alpha}_{s} K_{0}
(k_{1},k_{2}) \left( \log \frac{k^{2}_{1}}{k^{2}_{2}}\right)^{2} \, ,
\hspace{1cm} (s_{0} =k_{1}k_{2})\, .
\label{17}
\end{equation}

The eigenvalue function of this kernel is simply
\begin{equation}
\Delta (\alpha_{s}\chi_{1})= - \frac{\overline{\alpha}_{s}}{4} \chi_{0}''\, ,
\hspace{1cm} (s_{0} =k_{1}k_{2})\, .
\label{18}
\end{equation}
where $\chi_{0}=2\psi(1)+\psi(\gamma)-\psi(1-\gamma)$ is the leading eigenvalue. By adding
it to the ``irreducible'' contribution previously derived \cite{9}, we obtain
the total gluonic eigenvalue for the scale $s_{0} =k_{1}k_{2}$:
\begin{eqnarray}
\alpha_{s}\chi_{1}^{(g)} 
&=& \frac{\overline{\alpha}_{s}}{4} \left[ -\frac{11}{6}
	(\chi^{2}_{0}+\chi^{\prime}_{0})+\left(\frac{67}{9} -\frac{\pi^{2}}{3}\right) 
	\chi_{0}(\gamma) +6\zeta(3)
	\right.\\
&-& \left.
	\left(\frac{\pi}{\sin\pi\gamma}\right)^{2}
	\frac{\cos \pi\gamma}{3(1-2\gamma)} 
	\left( 11+\frac{\gamma(1-\gamma)}{(1+2\gamma)(3-2\gamma)}\right)
+	\left(\frac{\pi^{2}}{3\gamma(1-\gamma)} + \tilde{h}(\gamma)\right)
	-\chi_{0}'' \right] \, . \nonumber 
\label{19}
\end{eqnarray}

The final result in Eq~(19) agrees \footnote{Therefore, the comparison with
Ref.\cite{9} made after Eqs~(7) and (24) of Ref.(12) is misleading (preprint
version hep-ph/980290).} with the one in Eq~(14) of Ref.(12)
after the identification
\begin{equation}
\frac{\pi^{2}}{3\gamma(1-\gamma)} + \tilde{h} (\gamma) =
\frac{\pi^{3}}{\sin \pi\gamma} - 4 \Phi (\gamma) \, ,
\label{20}
\end{equation}
where in the l.h.s. (r.h.s.) we use the notation of Ref.(9) (Ref~(12)).

Since $\chi_{0} \sim 1/\gamma +2\zeta(3) \gamma^{2}+$...
for $\gamma\rightarrow0$, the additional term (18) has a nasty
$1/\gamma^{3} (1/(1-\gamma)^{3})$ singularity for $\gamma=0 (\gamma=1)$,
thus confirming that the scale $k_{1}k_{2}$ is not collinear safe for
$k_{1}\gg k_{2} (k_{2}\gg k_{1})$. However, one can change scale to
$k_{1}^{2}(k_{2}^{2})$ in Eq~(1) by a simple trick. The gluon Green's
function (actually, its scale-invariant part) has the representation
\begin{equation}
\int \frac{d\gamma}{2\pi i}\, \frac{d\omega}{2\pi i}
\left(\frac{s}{k_{1}k_{2}}\right)^{\omega}
g_{\omega}(\gamma) \frac{1}{k^{2}_{1}}
	\left(\frac{k_{1}^{2}}{k_{2}^{2}}\right)^{\gamma},
g_{\omega}(\gamma)=	\frac{1+\overline{\alpha}_{s} (h(\gamma)+h(1-\gamma))}
	{1-\frac{\overline{\alpha}_{s}}{\omega} 
	\left(\chi_{0}(\gamma)+\alpha_{s} \chi_{1}(\gamma)\right)}
	 \; .
\label{21}
\end{equation}
Therefore, a scale change $k_{1}k_{2} \rightarrow k^{2}_{1}(k^{2}_{2})$
is equivalent to replacing $\gamma$ with $\gamma +\frac{1}{2}\omega
(\gamma - \frac{1}{2}\omega)$. By shifting $\gamma$ and then expanding
in $\omega$ up to $NL$ level, it is straightforward to rewrite Eq~(21)
for the scale $\mathbf{k}^{2}_{1}$, say, with the Green's function
\begin{equation}
g^{(1)}_{\omega}(\gamma)
=	\frac{1+\overline{\alpha}_{s} \left(h(\gamma)+h(1-\gamma)
	-\frac{1}{2}\chi_{0}^{\prime}(\gamma)\right)}
	{1-\frac{\overline{\alpha}_{s}}{\omega} 
	\left(\chi_{0}(\gamma)+\alpha_{s} \chi_{1}-\frac{1}{2}
	\overline{\alpha}_{s}\chi_{0}\chi_{0}^{\prime}\right)}\, , 
	\hspace{1cm}
	(s_{0}=\mathbf{k}^{2}_{1})\, .
\label{22}
\end{equation}

This means that the scale-dependent contribution to the $NL$ kernel is,
for the scale $\mathbf{k}_{1}^{2}(\mathbf{k}^{2}_{2})$, given by
\begin{eqnarray}
\Delta (\alpha_{s}\chi_{1})
&=&	-\frac{\overline{\alpha}_{s}}{4} (\chi_{0}^{\prime} +
	\chi_{0}^{2})^{\prime} \, , \;\;\;\; (s_{0}=k^{2}_{1})\, , \\
&=&	-\frac{\overline{\alpha}_{s}}{4} (\chi_{0}^{\prime} -
	\chi_{0}^{2})^{\prime} \, , \;\;\;\; (s_{0}=k^{2}_{2})\, .
\nonumber
\label{23}
\end{eqnarray}
It is apparent that $\Delta\chi_{1}$ at scale $k_{1}^{2}$ is now perfectly
regular for $\gamma=0$, the $\gamma^{-3}$ singularity being cancelled,
while it still keeps the nasty cubic singularity at $\gamma =1$. 
Furthermore, by Eq.~(8), the numerator in Eq.~(22) becomes \textit{regular}
at $\gamma=0$ too, where it takes the value $(1+\overline{\alpha}_{s}
h(1))=(1-\overline{\alpha}_{s}\psi^{\prime}(1))$, which renormalizes the
impact factors, as noticed already in Ref~\cite{11}. This confirms the
importance of factorizing $H$ in the gluon Green's function in Eq.~(2).

In other words, the scale $k^{2}_{1}(k^{2}_{2})$ is proved to be
collinear safe for $k^{2}_{1}\gg k^{2}_{2} (k^{2}_{1} \ll k^{2}_{2})$,
but both cases are described by quite asymmetrical kernels, related by
an $\omega$-dependent similarity transformation to the symmetrical one
describing the scale $k_{1}k_{2}$.

In the anomalous dimension regime $k^{2}_{1} \gg k^{2}_{2}$, the
scale-dependent contribution in Eq~(23) changes the small $\gamma$
expansion of $\chi_{1}=A_{1}/\gamma^{2} + A_{2}/\gamma +A_{3}+$ ...
by just the constant $-2\zeta (3)$. Therefore the constant $A_{3}$
takes the value
\begin{equation}
A_{3} = \frac{N_{C}}{4\pi} \left( -2\zeta (3) - \frac{395}{27}
+ \frac{11}{18} \pi^{2} + \frac{\pi^{2}}{3} + \tilde{h}(0) \right)
\label{24}
\end{equation}
and determines the $NL$ anomalous dimension at 3-loop level. The
result (24) agrees with Eq~(23) of Ref~(12), because $\tilde{h} (0)
= 4\zeta (3) -\pi^{2}/3$.

Since $\chi^{\prime}_{0}(\frac{1}{2})=0$, the hard Pomeron shift,
as singularity of the anomalous dimension expansion \cite{8}, is independent of
which scale is being used in the factorized class $(k_{1}k_{2}$, or
$k^{2}_{1}$, or $k^{2}_{2}$). However, since $\chi_{0} ''(\frac{1}{2})$
is large, the shift of Ref~\cite{9} is strongly affected by
$\Delta \chi_{1}$ and becomes
\begin{equation}
\omega_{P}
=	\overline{\alpha}_{s} \chi_{0} (\frac{1}{2}) [1-(3.4+3.0)
	\overline{\alpha}_{s}]\; 
=	\overline{\alpha}_{s} \chi_{0} (\frac{1}{2}) [1-6.4
	\overline{\alpha}_{s}] \; , 
\label{25}
\end{equation}
so that the $NL$ correction is almost doubled.

Therefore, $\omega_{P}$ saturates at the small value 
$\omega_{P}^{\textrm{Max}}=.11$, \footnote{The value .214 quoted in Ref~(12)
(version hep-ph 980290) is
presumably misprinted by a factor of 2.} and for a very small value of
$\alpha_{s}=.08$. This means that the trend of a negative correction is
confirmed but, unfortunately, the next-to-leading hierarchy is even less
convergent than previously thought. 

What about switching to the (non-factorized)
scale $s_{M} =$ Max $(k^{2}_{1},k^{2}_{2})$? In this case the argument
after Eq~(21) applies with a $\gamma$-shift of $-\frac{1}{2} \omega
(+\frac{1}{2} \omega)$ according to whether $k_{1}>k_{2} (k_{1}<k_{2})$.
More precisely, the Green's function for scale $s_{M}$ is given by
\begin{equation}
\left. g^{(M)}_{\omega}(\gamma) \right]_{L} =
\left. g_{\omega}(\gamma -\frac{1}{2}\omega) \right]_{L} \, , \;\;
\left. g^{(M)}_{\omega}(\gamma) \right]_{R} =
\left. g_{\omega}(\gamma +\frac{1}{2}\omega) \right]_{R}
\label{26}
\end{equation}
where we have decomposed $g_{\omega}(\gamma)$ into its projections
$g_{L}(g_{R})$ having l.h. (r.h.) singularities in the $\gamma$-plane.

A careful analysis shows that the transformation (26) implies changing
the $NL$ quantities, at two-loop level, as follows
\begin{eqnarray}
h^{(M)}(\gamma)		
&=&	\left. h(\gamma)-\frac{1}{2} \chi^{\prime}_{0}\right]_{L} \, , 
	\hspace{1cm} s_{M}=\textrm{Max} (k^{2}_{1},k^{2}_{2}))\, , \\
\alpha_{s}\chi^{(M)}_{1}(\gamma)
&=&	\alpha_{s}\chi_{1}(\gamma) + \overline{\alpha}_{s}
	\left(\frac{1}{2} \chi_{0} [\chi^{\prime}_{0}]_{s} -
	[\chi_{0}\chi^{\prime}_{0}]_{s}\right) \, , \nonumber
\label{27}
\end{eqnarray}
where $O(\gamma)]_{s} =O(\gamma)]_{L} - O(\gamma)]_{R}$ is a symmetrical
combination of $L$ and $R$ projections. However, since such
projections act on \textit{non linear} combinations of $H$ and $K_{1}$, 
reproducing Eq~(26)
to all orders requires modifying the two-loop relation (27) by additional
powers of $\frac{\overline{\alpha}_{s}}{\omega}$ at higher orders.

Note also that the hard Pomeron shift implied by Eq~(27) is different from
 the result (25), due to a nonvanishing contribution
of the term in round brackets.
This somewhat puzzling feature is due to the fact just mentioned that
equivalent results in the two scales cannot be simply hinted at by a
two-loop calculation, but require reshuffling the whole perturbative series.
In other words, if $\chi_{1}$ iterates, $\chi_{1}^{(M)}$ in Eq.~(27) does
not, and viceversa.

To sum up, we have proved in this note the form (19) of the $NL$ eigenvalue
for the scale $k_{1}k_{2}$, by deriving the scale-dependent contributions
(18), (23) and (27). In the class of factorized scales, assuming 
Eq.~(1) to all orders, the $NL$ corrections are even larger than
previously thought. It becomes then important to supplement the $NL$ hierarchy
with finite-$x$ effects, perhaps along the lines suggested by the CCFM 
equation \cite{13,16}.

\end{document}